\documentclass{emulateapj}

\usepackage{placeins}

\shorttitle{Evolution of disks} \shortauthors{Zhu, Hartmann, \& Gammie}

\begin{document}

\title{Long-term Evolution of Protostellar and Protoplanetary Disks. I. Outbursts}

\author{Zhaohuan Zhu\altaffilmark{1}, Lee Hartmann\altaffilmark{1},
Charles F. Gammie \altaffilmark{2,3}, Laura G. Book\altaffilmark{4},
Jacob B. Simon\altaffilmark{5}, Eric Engelhard\altaffilmark{6} }

\altaffiltext{1}{Dept. of Astronomy, University of Michigan, 500
Church St., Ann Arbor, MI 48105} \altaffiltext{2}{Dept. of
Astronomy, University of Illinois Urbana-Champaign, 1002 W. Green
St., Urbana, IL 61801} \altaffiltext{3}{Dept. of Physics, University
of Illinois Urbana-Champaign} \altaffiltext{4}{Dept. of Physics,
California Institute of Technology, Pasadena, CA 91125}
\altaffiltext{5}{Dept. of Astronomy, University of Virginia, VA
22904-4325}
\altaffiltext{6}{G-Mart Comics}

\email{zhuzh@umich.edu, lhartm@umich.edu, gammie@illinois.edu}

\newcommand\msun{\rm M_{\odot}}
\newcommand\lsun{\rm L_{\odot}}
\newcommand\msunyr{\rm M_{\odot}\,yr^{-1}}
\newcommand\be{\begin{equation}}
\newcommand\en{\end{equation}}
\newcommand\cm{\rm cm}
\newcommand\kms{\rm{\, km \, s^{-1}}}
\newcommand\K{\rm K}
\newcommand\etal{{\rm et al}.\ }
\newcommand\sd{\partial}
\newcommand\mdot{\rm \dot{M}}
\newcommand\rsun{\rm R_{\odot}}
\newcommand\yr{\rm yr}

\begin{abstract}
As an initial investigation into the 
long-term evolution of protostellar
disks, we explore the conditions required to explain the large
outbursts of disk accretion seen in some young stellar objects. We
use one-dimensional time-dependent disk models with a phenomenological
treatment 
 of the magnetorotational instability (MRI) and
gravitational torques to follow disk evolution over long timescales.
Comparison with our previous two-dimensional disk model calculations
(Zhu et al. 2009b, Z2009b) indicates that the neglect of radial
effects and two-dimensional disk structure in the one-dimensional
case makes only modest differences in the results; this allows us to
use the simpler models to explore parameter space efficiently. We
find that the 
 mass infall rates typically estimated for low-mass
protostars generally result in AU-scale disk accretion outbursts,
 as predicted by our previous analysis (Zhu \etal 2009a,
Z2009a). We also confirm quasi-steady accretion behavior for high
mass infall rates if the values of $\alpha$-parameter for the
magnetorotational instability is small, while at this high accretion
rate convection from the thermal instability may lead to some
variations. We further constrain the combinations of the
$\alpha$-parameter and the MRI critical temperature, which can
reproduce observed outburst behavior. Our results suggest that dust
sublimation may be connected with full activation of the MRI. This
is consistent with the idea that small dust captures ions and
electrons to suppress the MRI.  In a later paper we will explore
both long-term outburst and disk
 evolution with this model, allowing
for infall from protostellar envelopes with differing angular
momenta.
\end{abstract}


\section{Introduction}
In the standard model of low-mass star formation, a molecular cloud
core collapses to a protostar over timescales of $\sim 10^{5}$~yr
(e.g., Shu, Adams, \& Lizano 1987), consistent with observations
(Kenyon \etal 1990; Enoch \etal 2008). However, steady accretion of
this mass onto central stars with a plausible mass-radius relation
results in accretion luminosities that are larger than those
observed in low-mass protostars (Kenyon \etal 1990, 1994; Enoch
\etal 2009). One solution to this ``luminosity problem'' is that
most infalling matter first falls to the circumstellar disk and then
is accreted to the star during short-lived outbursts; in this model
protostars are usually observed in quiescence. The FU Orionis
objects provide direct evidence for this type of behavior, with
maximum accretion rates of $10^{-4} \msunyr$ over periods of decades
to centuries (Hartmann \& Kenyon 1996), which also directly suggests
$\sim$10$^{-2}$ M$_{\odot}$ in the disk at $\sim$ 1 AU.

A number of theories have been proposed to explain FU Orionis
outbursts, including thermal instability in the inner disk (Bell
$\&$ Lin 1994), binary interactions \citep{Bonnel1992}, and
gravitational clumping at several AU (Vorobyov $\&$ Basu 2005,
2006).  In a recent paper (Z2009a), we explored the possibility that
outbursts might result because of a mismatch between the mass fluxes
that can be transported by the magnetorotational instability (MRI)
in the inner disk, and the gravitational instability (GI) in the
outer disk. Using steady thin disk theory, we argued that outbursts
are to be expected when disks are driven by mass addition lower than
10$^{-4}\msunyr$, as initially found by Armitage, Livio, \& Pringle
(2001). We then developed a two-dimensional model of FU Orionis
disks, which verified that outbursts of accretion similar to those
observed could be produced using reasonable parameters for MRI
transport when thermal ionization dominates (Z2009 b).

The computationally-intensive nature of two-dimensional (let
alone three-dimensional) simulations of outbursts makes it
difficult to conduct studies of the effects of differing
parameters on disk evolution over significant timescales.  We have
therefore developed one-dimensional disk models to follow disk
evolution.  While such models have limitations, they can
serve as a starting point to investigate the landscape of possible
disk evolutionary pathways. The diversity of disk properties among
stars of nearly the same age and mass (e.g., Hartmann 2009)
is plausibly
 the result of differing initial conditions, and models
of the type we explore here can begin to address this possibility.

In \S 2 we describe our 1-D, two-zone model, and compare its
outburst properties with our 2-D models in \S 3.  In \S 4 we present
the results of a parameter study designed to show when outbursts
occur and how the outburst strength and frequency depend on the
adopted parameters.  We present our discussion and conclusions in \S
5 and \S 6.  The present work, in which we assume constant mass
addition to the outer disk, serves as a starting point for our
subsequent investigation of long-term disk evolution with mass
addition due to infall from a rotating protostellar envelope in a
following paper.

\section{One-dimensional two-zone models}

In this paper we adopt a version of the ``layered accretion'' disk
model originally put forward by Gammie (1996). In this model, unless
the disk is warm enough that thermal ionization is sufficient to
couple the magnetic field effectively to the neutral disk material,
only an upper layer of the disk can sustain the MRI due to
non-thermal ionization by cosmic and/or X-rays. A significant amount
of work has been done to study the properties of the active layer
(e.g. Sano \etal 2000, Turner \& Sano 2008, Bai \& Goodman 2009),
however due to the complex physics and chemistry involved, further
theoretical or even observation study is needed. Here, we assume
that the mass column density which can be ionized is roughly
constant with radius (see, e.g., Glassgold, Najita \& Igea
 2004). If the total disk surface density $\Sigma$ in the
``cold'' regions is less than the limiting active layer column
density $\Sigma_a$, the disk is assumed to be completely viscous
with a given $\alpha_M$ viscosity parameter due to MRI activity. On
the other hand, if the disk surface density is higher than this
limit, then only the surface layers are assumed to exhibit the MRI.

Our one-dimensional, two-zone models (1D2Z) thus in general exhibit
two layers: the surface layer and the central or ``dead'' zone
\footnote{Here "dead" refers to magnetically dead without MRI but it
could transport angular momentum due to the gravitational
instability.} with surface density $\Sigma_d$. The dead zone is
assumed to have little or no MRI activity, though it may exhibit
transport due to GI (see below). The temperatures T$_{a}$ and
T$_{d}$ are averages which characterize the corresponding layers.

The surface density evolves
according to the mass conservation and angular momentum conservation
equations,
\begin{equation}
2\pi R\frac{\partial \Sigma_{i}}{\partial t}=\frac{\partial
\dot{M_{i}}}{\partial R}\,,\label{eq:1}
\end{equation}
\begin{equation}
2\pi R\frac{\partial}{\partial t}(\Sigma_{i} R^{2}
\Omega)=\frac{\partial}{\partial
R}(\dot{M_{i}}R^{2}\Omega)+\frac{\partial}{\partial R}(2\pi
R^{2}W_{R\phi,i})\,,\label{eq:2}
\end{equation}
where $\dot{M_{i}}=-2\pi\Sigma_{i} R v_{i}$ is the radial mass flux
in the disk, the stress $W_{R\phi,i}=R\Sigma_{i} \nu_{i}d\Omega/dR$,
and subscript $i$ denotes either 'a' (active layer) or 'd' (dead
zone).

Assuming Keplerian rotation, equations (\ref{eq:1}) and (\ref{eq:2})
can be simplified to
\begin{equation}
\partial_{t}\Sigma_{i}=\frac{2}{R}\partial_{R}(\frac{1}{R\Omega}\partial_{R}(-R^{2}W_{r\phi}))\,,\label{eq:3}
\end{equation}
where $\partial_{R}\equiv\partial/\partial R$. The constant infall
rate $\dot{M_{in}}$ is set as an inflow outer boundary condition
with $\dot{M_{in}}$=10$^{-4}$, 10$^{-5}$, and 10$^{-6}$$\msunyr$.

The parameter $\Sigma_{A}$ is the maximum non-thermally ionized
surface density, assumed constant during each calculation. If at a
given timestep $\Sigma_{a}>\Sigma_{A}$, the excess mass of the
active layer ($\Sigma_{a}$-$\Sigma_{A}$) is added to the dead zone
($\Sigma_{d}$=$\Sigma_{d}$+{$\Sigma_{a}$-$\Sigma_{A}$}). Conversely,
if $\Sigma_{a}<\Sigma_{A}$ and $\Sigma_{d}\neq0$, part of the dead
zone is assumed to be non-thermally ionized, in which case
$\Sigma_{a}$ is set to be $\Sigma_{A}$ and $\Sigma_{d}$ decreases to
$\Sigma_{d}-(\Sigma_{A}-\Sigma_{a})$.  Setting $\Sigma_{A} = const.$
is a crude approximation; it is likely that $\Sigma_{A}$ varies with
radius and depends on the local abundance of dust and flux of
ionizing radiation.

The temperatures are determined by the balance between the heating and
radiative cooling,
\begin{equation}
C_{\Sigma,i}\partial_{t}T_{i}=Q_{heat,i}-Q_{cool,i}\,,
\end{equation}
where the heat capacity is $C_{\Sigma,i}$=$\Sigma_{i}
c_{s,i}^{2}/T_{i}$.
For the active layer, the cooling rate is determined by
\begin{equation}
Q_{cool,a}=\frac{16}{3}\sigma(T_{a}^{4}-T_{ext}^{4})\frac{\tau_{a}}{1+\tau_{a}^{2}}\,,\label{eq:qcool}
\end{equation}
where $\sigma$ is the Stefan-Boltzmann constant, $T_{ext}$
represents the irradiation from the central star, and $\tau_{a}$ is
the optical depth of the active layer. The final factor in equation
(\ref{eq:qcool}) is an approximate form which accommodates both
optically thin and thick cooling.  We assume
\begin{equation}
T_{ext}^{4}=fL /(4\pi R^{2}\sigma)\,,\label{eq:text}
\end{equation}
where $L$ is the total luminosity of the star and $f(R)$
accounts for the non-normal irradiation of the disk by the central
star; here we set $f(R) = const. = 0.1$.
The active layer optical depth is given by
\begin{equation}
 \tau_{a}=\frac{1}{2} \Sigma_{a} \kappa(\rho_{a},T_{a})\,,  \label{eq:tau}
\end{equation}
where $\kappa$ is the Rosseland opacity derived from Z2009a at the
active layer density and temperature, $\rho_{a}=\Sigma_{a}/2H_{a}$,
$\rho_{d}=(\Sigma_{a}+\Sigma_{d})/2H_{d}$, and $H_{a}$ and $H_{d}$
are the scale height of the active layer and the dead zone.

The dead zone has a cooling rate similar to that of the active layer. If the
active layer is optically thick, the incident radiation flux into
the dead zone from the active layer is $\sigma$T$_{a}^{4}$. Thus the
cooling rate is
\begin{equation}
Q_{cool,d}=\frac{16}{3}\sigma(T_{d}^{4}-T_{a}^{4})\frac{\tau_{d}}{1+\tau_{d}^{2}}\,,
\end{equation}
with
\begin{equation}
 \tau_{d}=\frac{1}{2} \Sigma_{d} \kappa(\rho_{d},T_{d})\,.  \label{eq:tau2}
\end{equation}
However, if the active layer is optically thin, the incident flux
becomes $\sigma$(T$_{a}^{4}$$\tau_{a}$+T$_{ext}^{4}$), and
\begin{equation}
Q_{cool,d}=\frac{16}{3}\sigma(T_{d}^{4}-T_{a}^{4}\tau_{a}-T_{ext}^{4})\frac{\tau_{d}}{1+\tau_{d}^{2}}\,.
\end{equation}
With increasing temperature the radiative cooling time, which
constrains the numerical timestep, becomes small.  For
computational efficiency we then make the equilibrium approximation
\begin{equation}
T_{a}^{4}=\frac{8(T_{d}^{4}-T_{ext}^{4})}{3\tau_{d}}\,,
\end{equation}
when the disk midplane temperature is $> 5000$K.

The heating rate of the dead zone is just the viscous heating rate,
while, in the active layer, the heating rate consists of its own
viscous heating and the radiation from the underlying dead zone:
\begin{equation}
Q_{heat,a}=Q_{visc}+Q_{cool,d}\,,
\end{equation}
In the case when the active layer is optically thin, this equation
is modified to
\begin{equation}
Q_{heat,a}=Q_{visc}+\tau_a Q_{cool,d}\,.
\end{equation}

The viscous heating term is
\begin{equation}
Q_{visc}=\frac{3}{2}W_{R\phi}\Omega\,,
\end{equation}
where $W_{R\phi}=(3/2)\Sigma_i \nu_{i}\Omega$. To evaluate the
viscosity $\nu_i$, we have considered both MRI and GI transport. The
net viscosity $\nu_i$ is the sum of both,
\begin{equation}
\nu_i=\alpha_i\frac{c_{s_i}^{2}}{\Omega}
\end{equation}
where $\alpha_i=\alpha_{Q}+\alpha_{M}$ and
\begin{equation}
\alpha_{Q}=e^{-Q^{4}}\,. \label{eq:Q}
\end{equation}
The MRI viscosity is assumed to have a fixed value of $\alpha_M$
whether the region in question is thermally or non-thermally
ionized.  We assume that above some critical temperature $T_{M}$ the
MRI is fully activated throughout the disk with viscosity parameter
$\alpha_{M}$. The Toomre instability parameter $Q$ is evaluated
using the disk central (midplane) temperature T$_{d}$, the total
surface density ($\Sigma_{a}$+$\Sigma_{d}$), and assuming Keplerian
rotation. The form of $\alpha_Q$ is motivated by a desire to make
gravitational torques significant only when $Q \lesssim 1.4$, as
indicated by global three-dimensional simulations (e.g., Boley et al
2006).

There are uncertainties in adopting the above approach to transport.
GIs involve large scale density waves that cannot be captured with a
local viscous treatment, although local treatments are adequate under
some circumstances \citep{Lodato2004,Cossins2009}.  As discussed in
Z2009a, the essential properties of this treatment are the assumptions
that disks with GI have $Q$-values of order unity, and that the GI
produce local dissipation of the accretion energy.  Under these
assumptions, the precise form of $\alpha_{Q}$ will not affect the disk's
evolution, as long as $\alpha$ is a steeply declining function of Q near
$\sim$1.5. To make this point clear, we ran the same simulation for a
test case but with the $\alpha_{Q}$ prescription of Lin \& Pringle
(1987,1990). As expected, the different forms of $\alpha_{Q}$ have no
effect on the disk outbursts.

Similarly, whether the MRI can be fully activated in a
non-thermally-ionized layer depends in part upon whether small dust
grains have been sufficiently depleted (e.g., Sano \etal 2000). This
is a complicated problem with substantial observational and
theoretical uncertainties; we therefore adopt the simplest possible
approach.  It turns out that the value of $\Sigma_a$ is unimportant
for understanding large outbursts (Z2009a), as long as the GI in the
dead zone transports more mass than the active layer; but
$\Sigma_{a}$ does have important effects on the long-term disk
evolution at low accretion rates, as discussed in a following paper.

It is now clear that the magnetic fields that give rise to
$\alpha_M$ diffuse radially (Lesur \& Longaretti 2008, Guan \&
Gammie 2009, Fromang \& Stone 2009) and take time to build up and
decay (e.g.  Hirose et al. 2009). To account for these effects we
introduce an evolution equation for $\alpha_M$:
\begin{equation}
\frac{\partial \alpha_M}{\partial t} = -\Omega \frac{\alpha_M^{2} -
\alpha_{M,o}^{2}}{\alpha_M^{2} + \alpha_{M,o}^{2}} +
\frac{H^{2}\Omega}{2}\frac{\partial^{2} \alpha_M}{\partial R^{2}}\,,
\end{equation}
where $\alpha_{M,o}$ is the equilibrium value for $\alpha_M$.  The first
term permits $\alpha_M$ to relax up, or down, as $T_M$ is crossed.   The
second term corresponds to radial diffusion of the magnetic field.  For
numerical reasons we set the dimensionless radial diffusion coefficient
to $0.5$ (the radial diffusion coefficient is actually a function of
distance from the midplane).

\section{Outburst behavior}

In the protostellar phase, the disk is unlikely to transport mass
steadily from $\sim$100 AU all the way to the star at an accretion
rate matching the mass infall rate $10^{-6}-10^{-4}\msunyr$ from the
envelope to the outer disk. This mismatch leads to outbursts which
are qualitatively similar to that found by \cite{armitage01}, Book
\& Hartmann (2005), and in our 2-D hydrodynamic simulations
(Z2009b). In summary, before the outburst, mass added to the outer
disk moves inwards due to GI, but piles up in the inner disk as GI
becomes less effective at smaller radii. Eventually, the large
$\Sigma$ and energy dissipation leads to enough thermal ionization
to trigger the MRI at several AU. The MRI front quickly moves in
across the inner disk and the inner disk accretes at a higher mass
accretion rate, resembling FU Orionis-type outbursts. This high mass
accretion rate during the outburst also makes the inner disk
thermally unstable. After the inner disk has been drained by the
outburst and becomes too cold to sustain the MRI, the disk returns
to the low state. With the mass continuously accreted from the outer
radii (or from an infalling envelope), the disk evolves to
conditions leading to another outburst. This MRI triggered by GI
outburst can also be understood as the classical thermal
instability, but the S-curve is formed primarily by the variation of
$\alpha$ near T$_{M}$, rather than variations in opacity and assumed
variation in $\alpha$ near hydrogen ionization (e.g. \cite{bell94}).

To test how 1D2Z models simulate the outbursts compared with 2-D
simulations, we set up a test case with all the parameters adopted
from our previous 2-D simulations (Z2009b). In both 1-D and 2-D
simulations, we have used an updated opacity from Z2009a and
T$_{M}$=1500 K . Because Z2009b do not consider irradiation,
the irradiation factor f in equation (\ref{eq:text}) for the 1D2Z
simulation is set to be 0.  The inner radius in both cases is
set to 0.2 AU.

\begin{figure}
\epsscale{.80} \plotone{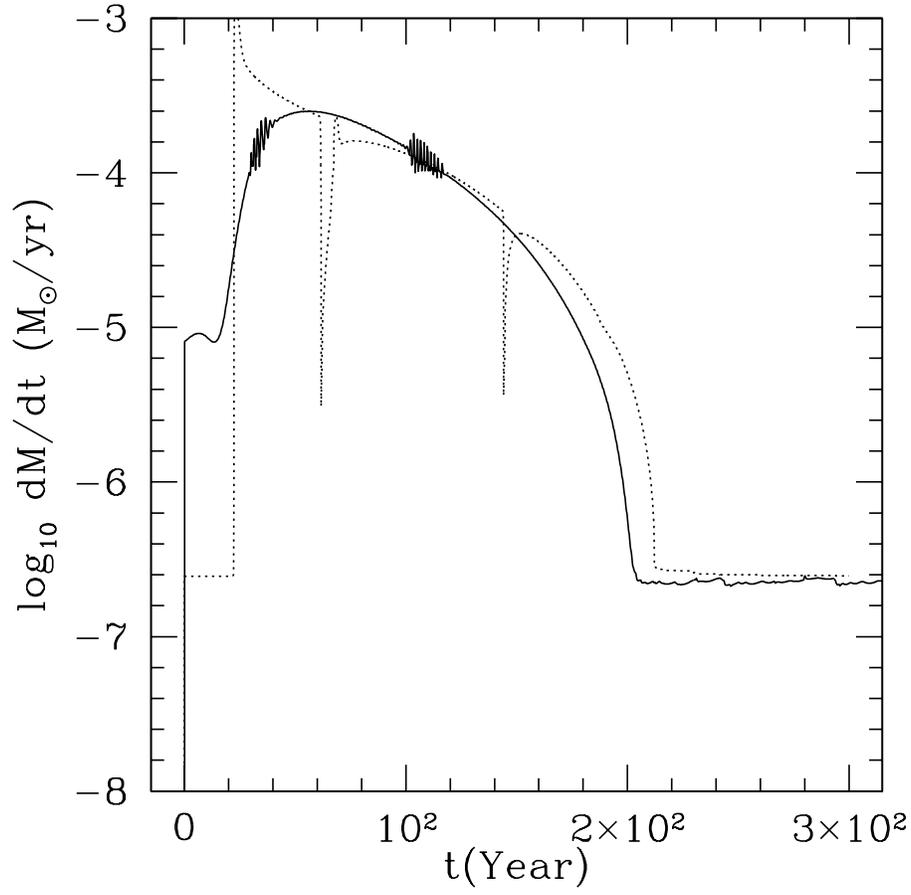} \caption{The mass accretion rate
with time for both 1-D (dotted curve) and 2-D (solid curve)
simulations.} \label{fig:dmcompare}
\end{figure}

Figure \ref{fig:dmcompare} shows the mass accretion rate as a function
of time for both 1-D and 2-D simulations. As shown, the 1-D simulations
closely resemble 2-D simulations at the equilibrium states, such as the
state before the outburst is triggered and the state during the
outburst.  For some rapid, or small scale, disk variations, such as the
MRI front propagation and the convective eddies in the hot inner disks,
the 2-D simulations exhibit more complex behavior than the 1-D
simulations, so the outbursts differ in detail.  In particular the 1-D
simulations show an initial high $\dot{M}$ peak at the beginning of the
outburst that is not seen in 2-D.  The 1-D simulations also show ``drop
outs'' in accretion that do not occur in 2-D.

In detail, starting from the MRI activation at $\sim$2 AU, the MRI
active region moves inwards.  During this process, mass piles up at the
inner boundary of the active region, because the disk is MRI active
beyond this boundary and has a higher mass accretion rate than the MRI
inactive disk at smaller radius. In 1-D simulations, which do not
capture the effects of radial pressure gradients, mass piles up at this
boundary.  When this mass eventually accretes on to the central star
there is a sharp peak in $\dot{M}$.

The $\dot{M}$ drop-outs during outbursts in the 1-D simulations are
related to the thermal instability associated with hydrogen ionization.
The drop-outs occur when the inner disk returns to the TI low state in
1-D; in 2-D radial and vertical convection smooths out variations in the
accretion rate and allows the inner disk to remain in the high state.

Despite the initial $\dot{M}$ peak and the drop-outs in the 1-D model,
the outburst timescale, $\dot{M}$, and consequently the total mass
accreted during one outburst are similar in 1-D and 2-D simulations
(Fig. \ref{fig:dmcompare}). This similarity is due to the fact that the
outburst timescale and $\dot{M}$ are just determined by the radius where
the MRI is triggered and by $\alpha_{M}$; the total mass accreted during
one outburst is the mass difference before and after the outbursts,
which are both equilibrium states.

\section{Parameter study}

Having tested that the 1-D models reproduce the general properties
of the outbursts (maximum $\dot{M}$, duration time, etc.), we next
turn to a parameter study with 1-D models to test the predictions of
Z2009a using steady state models.

The radial range considered is from $\sim$0.2 AU to 30 AU.  We adopt a
constant inflow boundary condition at 30 AU. The inner boundary 0.2-0.3
AU is chosen to avoid instabilities which occur at the smallest radius
where there is a transition between thermal and non-thermal activation
of the MRI, here called $R_i$.

We find that, at least for our phenomenological model, $R_i$ is
unstable. \cite{Wunsch} has also found even with a 2-D radiative
transfer hydrodynamic layered model, this instability still occurs.
$R_{i}$ oscillates around a mean radius, with thermal fronts washing
inward and outward. This is similar to thermal instability but due
to $\alpha$ variations between the dead zone and the inner MRI
active region \citep{Wunsch}. However, we should be cautious in this
instability, especially since dust may be sublimated before MRI
activation. Then the inner disk may become optically thin due to the
low opacity of the gas, and direct irradiation from the central star
would ionize the dust wall. A proper treatment of $R_i$ needs a 3-D
MHD simulation with irradiation, dust physics, and ionization
physics. Because this treatment is impractical, we set the inner
boundary of our model just outside $R_i$.  $R_i$ depends on a
variety of parameters (central star luminosity, the irradiation
angle to the disk's surface, active layer surface density, the MRI
trigger temperature and its viscosity parameter).  Generally, $R_i$
increases as the heating rate increases.  In the end we set the
inner boundary to $0.2$AU if $\alpha_{M}$=0.01 and 0.3 AU if
$\alpha_{M}$=0.1.

The infall is treated as a constant mass inflow at the disk outer
edge at 30 AU. This assumption may not be applicable to study the
disk's long-term evolution during the entire infall stage (10$^{5}$
years) since the infall centrifugal radius increases with time; this
is explored extensively in Paper II (Zhu \etal 2010b). Over the
timescale of a single outburst, as studied in this paper, this
assumption is a good approximation as long as the infall centrifugal
radius is larger than the MRI stable radius (R$_{M}$~1-10 AU in
Figure \ref{fig:m1}).

Because protostars are thought to form over $\sim$ few $\times
10^{5}$ yr, the average infall rate should be
10$^{-6}$-10$^{-5}\msunyr$ (Stahler 1988; Hartmann et al. 1997);
numerical simulations suggest the infall rate could be up to
10$^{-4}\msunyr$ at the earliest stages (Bate et al. 2003).  Thus we
study the disk evolution with infall rates varying from
10$^{-4}\msunyr$ to 10$^{-6}\msunyr$.

We neglect any change in the central star mass over the outburst
timescale, which is a reasonable approximation.  The effect of
changing central mass over longer evolutionary timescales is
included in Paper II.

What is $\alpha_M$?  Observations of dwarf novae, X-ray binaries,
and FU Ori suggest $\alpha \sim 0.1$ (King et al.  2007; Zhu et al.
2007).  T Tauri disk observations suggest $\alpha \sim 0.01$
(Hartmann et al. 1998).  MHD simulations suggest $\alpha \gtrsim
0.01$, with the precise value depending on the resolution, treatment
of small-scale dissipation, stratification, and treatment of
radiation transport (e.g. Fromang \& Papaloizou 2007, Guan et al.
2009, Davis et al. 2009, Shi et al. 2009; Hartmann et al. 1998).
Because the situation is not yet resolved, we will consider cases
with $\alpha_M = 0.01$ and $\alpha_M = 0.1$.

The MRI activation temperature $T_{M}$ is not known precisely (it
depends on the location of alkali metals, their ionization rate, the
abundance of small grains, and the threshold ionization fraction for
MRI turbulence), so we consider cases with $T_{M}$=1400 K and 1800
K.  These values are chosen, consistent with our opacity
prescription, to represent cases with and without dust.

\subsection{Dependence on $\dot{M_{in}}$}

Figures \ref{fig:1em4} -- \ref{fig:1em62}
 show the outbursts with
different infall rates varying from 10$^{-4}\msunyr$ to
10$^{-6}\msunyr$. For a given set of disk parameters ($\alpha_{M}$,
T$_{M}$), the disk accretes quasi-steadily if the infall rate is high
enough (small variations could still appear due to the classical thermal
instability, as discussed in \S 5.1).  If $\alpha_{M}=0.01$, the disk
accretes steadily if the infall rate is $10^{-4}\msunyr$, while
outbursts appear with smaller infall rates. With a bigger
$\alpha_{M}=0.1$, the disk accretes nonsteadily/in outburst for all the
cases with infall rates $10^{-4}\msunyr$-$10^{-6}\msunyr$. This is
consistent with our predictions from the steady state models (Z2009a).

The outbursts become shorter with smaller infall rates
$\dot{M_{in}}$. For example, with $\alpha_{M}$=0.1 and T$_{M}$=1400
K, the outbursts last 800, 500, and 400  years for
$\dot{M_{in}}=10^{-4}$, $10^{-5}$, and $10^{-6}
\msunyr$.
 This can be explained by the fact that the MRI is
triggered at larger radii with larger $\dot{M_{in}}$ and thus the
outburst (viscous) timescale is longer (\S5.1). The simple analytical
calculation in the Appendix shows that the outburst timescale is
\begin{equation}
t\sim\frac{R_{Q}^{2}}{\nu}\sim 960\frac{0.1}{\alpha_{M}}
\left(\frac{\dot{M_{in}}}{10^{-4}\msunyr}\right)^{1/9}\left(\frac{M}{M_{\odot}}\right)^{2/3}
\rm{yr}\,.\label{eq:timescale}
\end{equation}
for T$_{M}$=1400 K, where $R_{Q}$ is the GI-induced MRI activation
radius which will be discussed in \S 5. This agrees with the
numerical simulations reasonably well. The mass accretion rates
during the outbursts are similar.

The infall rate also determines the outburst frequency. The time between
two outbursts is
 significantly shorter with $\dot{M_{in}} =
10^{-4}\msunyr$ than with
$\dot{M_{in}}=10^{-5}\msunyr$, because the mass accreted to the star
during all the outbursts should be equal to the mass from infall
integrated over the same period of time.  Since the outburst
timescale is insensitive to the infall rate ($\sim
\dot{M_{in}}^{1/9}$), we assume each outburst transports $\sim 0.03
M_{\odot}$, which is suggested by the observation of FU Ori, and
thus the timescale between two outbursts is 0.03
M$_{\odot}$/$\dot{M_{in}}$. Therefore higher infall rates lead to
more frequent outbursts.

\subsection{Dependence on $\alpha_{M}$ and T$_{M}$}
\begin{figure}
\epsscale{.80} \plotone{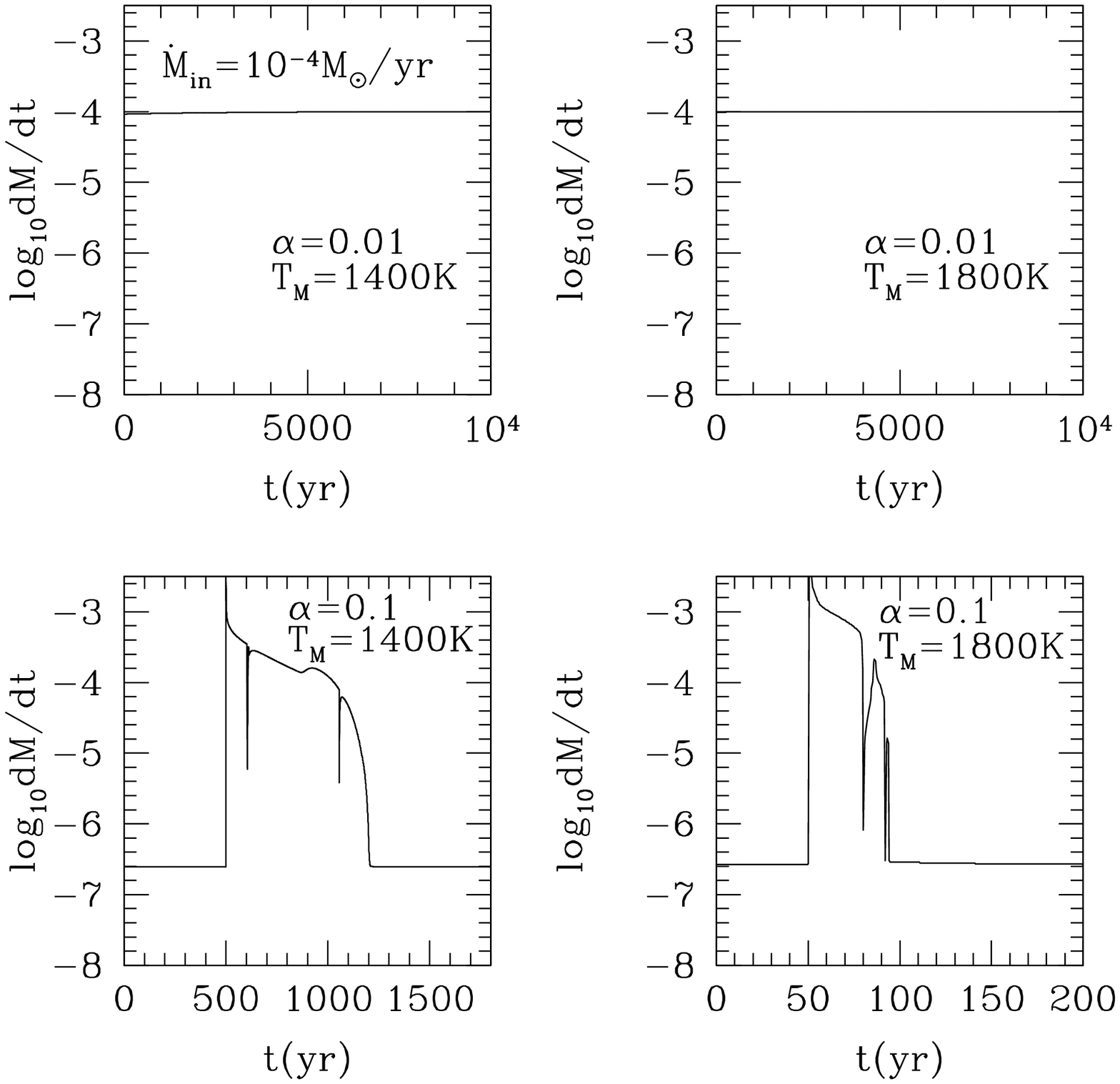} \caption{The mass accretion rate
with time for different $\alpha_{M}$ and T$_{M}$ for the mass infall
rate of 10$^{-4}$$\msunyr$.} \label{fig:1em4}
\end{figure}

\begin{figure}
\epsscale{.80} \plotone{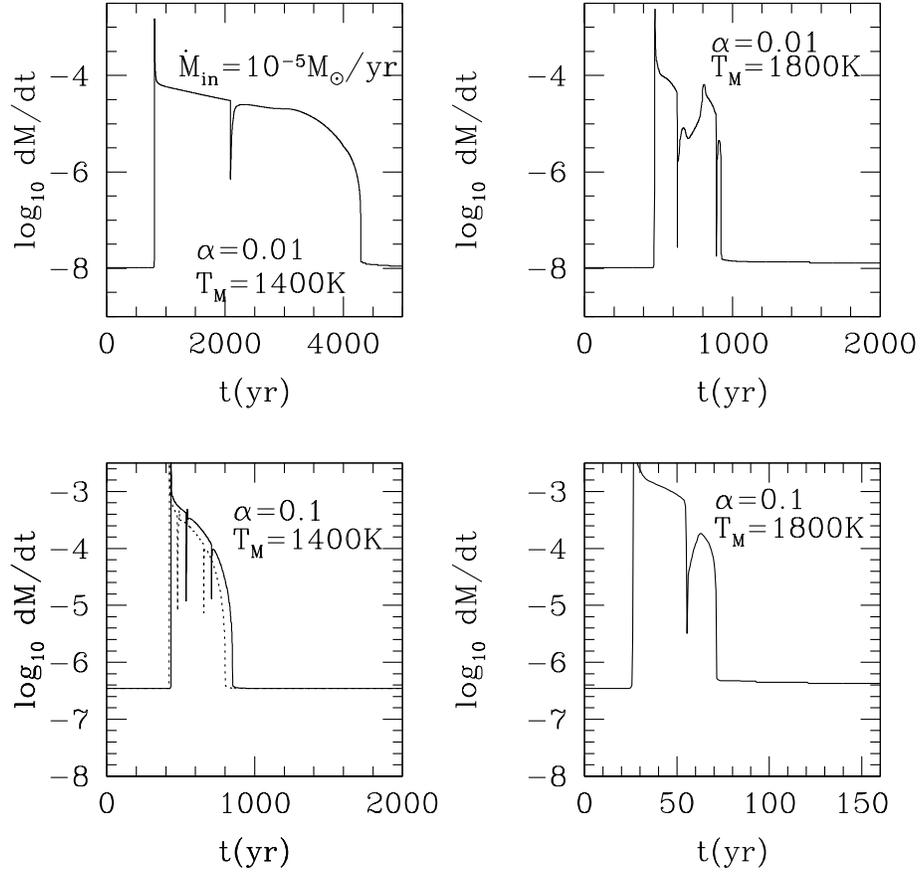} \caption{The mass accretion rate
with time for different $\alpha_{M}$ and T$_{M}$ for the mass infall
rate of 10$^{-5}$$\msunyr$. Compared with the solid curves which are
from the simulation with $\alpha_{Q}$=exp(-Q$^{4}$), the dotted
curve in the lower left panel shows the outburst obtained using the
$\alpha_{Q}$ prescription of Armitage \etal (2001). }
\label{fig:1em52}
\end{figure}

\begin{figure}
\epsscale{.80} \plotone{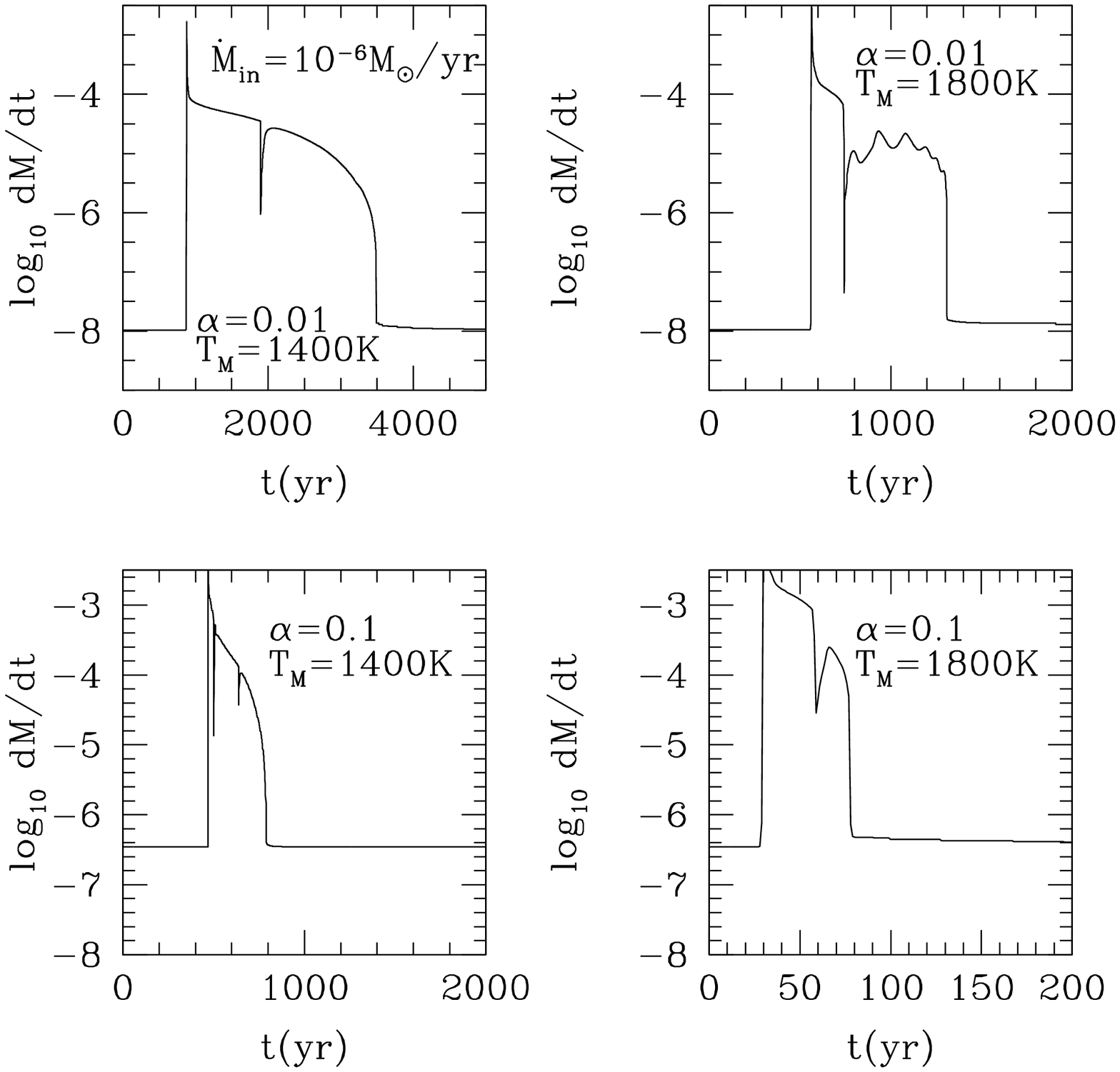} \caption{The mass accretion rate
with time for different $\alpha_{M}$ and T$_{M}$ for the mass infall
rate of 10$^{-6}$$\msunyr$.} \label{fig:1em62}
\end{figure}

The effect of $\alpha_{M}$ on the outburst can be seen by comparing
the upper and lower panels of
 Figures
\ref{fig:1em4}-\ref{fig:1em62}. For a given $\dot{M_{in}}$ and
$T_{M}$, with a higher $\alpha_{M}$ the outburst is shorter and
stronger. This is because the outburst timescale is close to the
viscous timescale, which is inversely proportional to $\alpha_{M}$
while the mass accretion rate is proportional to $\alpha_{M}$. This
can be understood using Equation (\ref{eq:timescale}) and
\begin{equation}
\dot{M}=5\times10^{-4}\frac{\alpha_{M}}{0.1}\msunyr\,,
\end{equation}
as shown in the Appendix for the $T_{M}$=1400 K case.

\begin{figure}
\epsscale{.80} \plotone{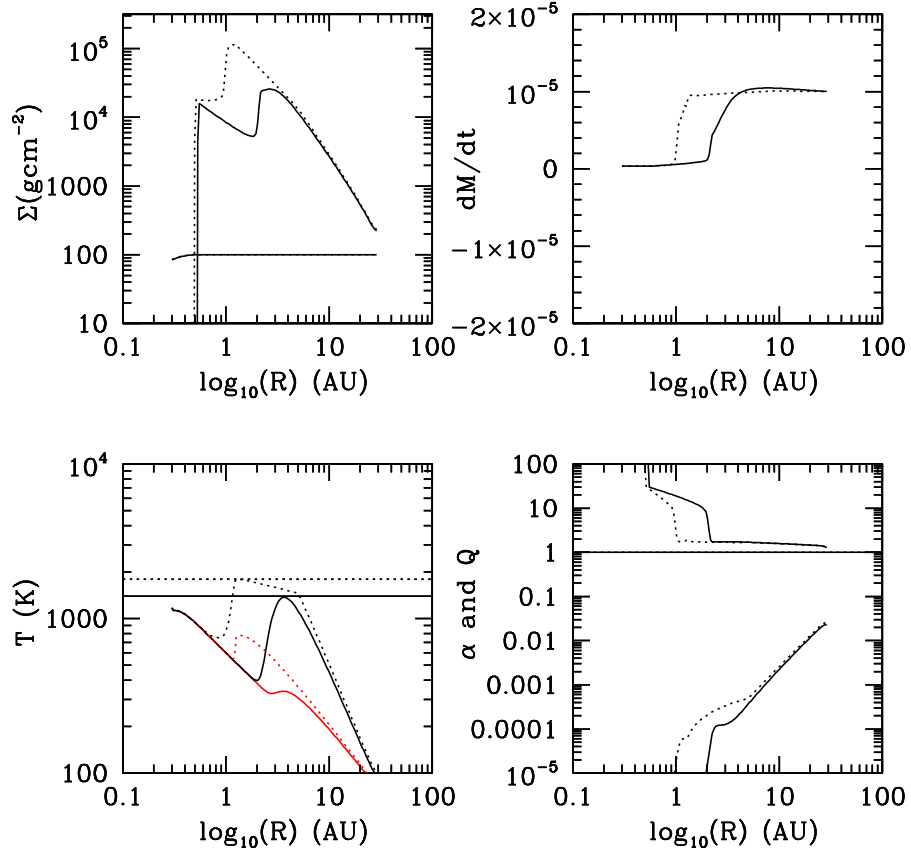} \caption{The disk's radial
structure at the stage just before the MRI is triggered in cases
where T$_{M}$=1800 K (dotted curve) and T$_{M}$=1400 K (solid
curve). The horizontal dotted line in the lower left panel shows
T$_{M}$=1400 K. The MRI is triggered at $\sim$ 3 AU for T$_{M}$=1800
K case and at $\sim$ 1.5 AU for T$_{M}$=1400 K case.}
\label{fig:14001800}
\end{figure}

On the other hand, comparing the left and right panels of Figures
\ref{fig:1em4} - \ref{fig:1em62},
 we find that the disks with
higher $T_{M}$ have shorter but stronger outbursts. The outburst is
shorter because the MRI is triggered at a smaller radius if $T_{M}$
is higher, resulting in a shorter viscous timescale. Figure
\ref{fig:14001800} shows the disk's condition just before the MRI is
triggered in the case $T_{M}$=1400 K (the solid curve) and
$T_{M}$=1800 K (the dotted curve). The MRI is triggered at 3 AU with
$T_{M}$=1400 K and 1.5 AU with $T_{M}$=1800 K. The outburst is
stronger with higher T$_{M}$ because the surface density at the MRI
trigger radius with $T_{M}$=1800 K (1 AU) is much higher than the
surface density at the MRI trigger radius with $T_{M}$=1400 K (3
AU) (upper left panel in Fig. \ref{fig:14001800}). Thus, with higher
T$_{M}$, the smaller but more massive inner disk leads to shorter
but stronger outbursts.

Our simulations also indicate
 that the outbursts with higher
$T_{M}$ accrete less mass than those with lower $T_{M}$ no matter
what the infall rate is (Table 1), because the outbursting inner
disk extends to smaller radii, and thus contains less mass, for a
higher $T_{M}$.

\subsection{Dependence on $M_{*}$ and $\Sigma_{A}$}

Protostars have a variety of masses, most of which are less than 1
M$_{\odot}$. We also calculated results for central star masses ( 0.3,
and 0.1 M$_{\odot}$) with the same T$_{M}$ (1400 K) and $\alpha_{M}$=0.1,
and the infall rate $\dot{M_{in}}=10^{-5}\msunyr$.
 The mass accretion
rates with time for these cases are shown in Figure \ref{fig:difmass}.
The outbursts have similar mass accretion rates, but the outburst is
slightly shorter with a less massive central star, as suggested by
equation (\ref{eq:timescale}). In addition the mass accreted during an
outburst is not significantly affected by its central star mass (Table
1). The outbursts are broadly similar, even though the central star
masses differ by a factor of 10.

\begin{figure}
\epsscale{.80} \plotone{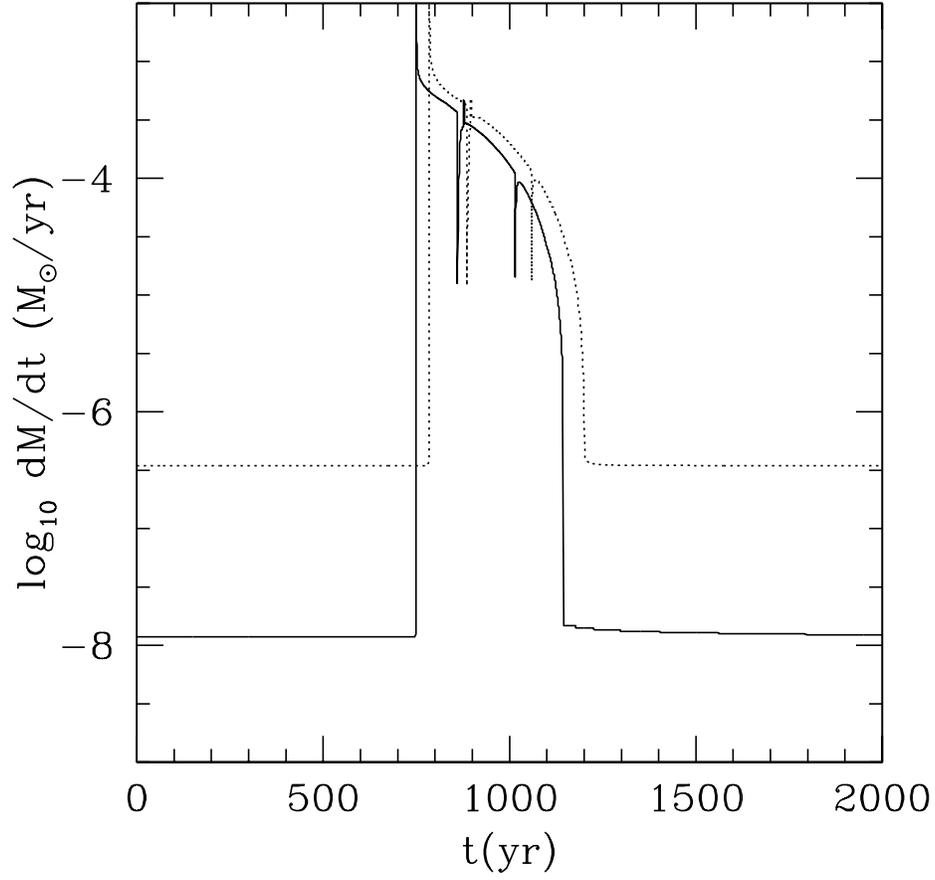} \caption{The disk mass accretion
rate with time for different active layer surface density (50 g
cm$^{-2}$ for the dotted curve and 10 g cm$^{-2}$ for the solid
curve). $\alpha_{M}$=0.1 and T$_{M}$=1400 K.  }
\label{fig:diftaunei}
\end{figure}

\begin{figure}
\epsscale{.80} \plotone{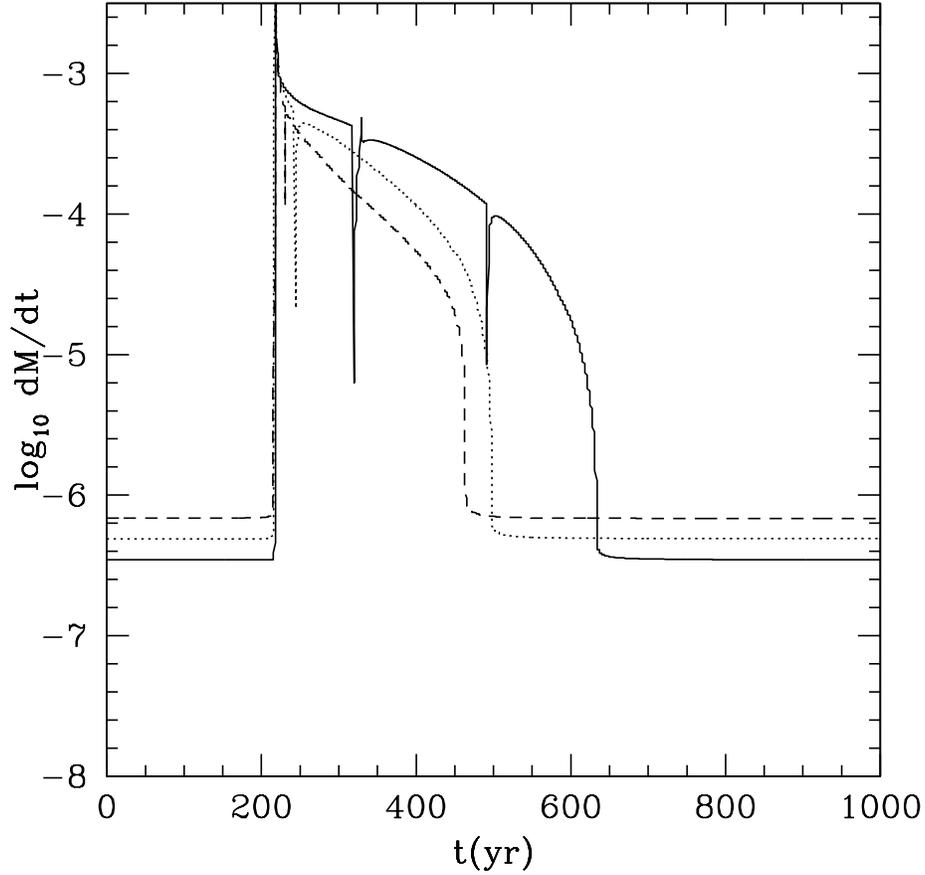} \caption{The mass accretion rate
with time for different central star masses: 1 M$_{\odot}$ (solid
curve), 0.3 M$_{\odot}$ (dotted curve), 0.1 M$_{\odot}$ (dashed
curve). The infall rate is 10$^{-5}$$\msunyr$, T$_{M}$=1400 K, and
$\alpha$=0.1.} \label{fig:difmass}
\end{figure}

Another uncertainty in a layered disk model is $\Sigma_{A}$, which
depends on the flux of ionizing radiation and the abundance and size
distribution of the dust.  In protostars with mass infall rates from
10$^{-6}\msunyr$ to 10$^{-4}\msunyr$, most of the infall mass is
transported to the inner disk by the GI in the dead zone, and mass
transport through the active layer is negligible.  Thus $\Sigma_{A}$
has little effect on the unsteady accretion and the outburst (Fig.
\ref{fig:diftaunei}), but it does affect the mass accretion rate in
the low state.  The mass accretion rate in a layered disk is
determined by the active layer mass accretion rate at $R_i$.  At $R
< R_i$ the disk is MRI active due to thermal ionization.  Thus the
disk mass accretion rate should be proportional to $\Sigma_{A}
\alpha_M$, which is shown in Figure \ref{fig:diftaunei}.

\section{Discussion}

\subsection{Unsteady accretion region}

The above parameter study can be simply summarized in the $\mdot - R$
plane as shown in Figure \ref{fig:m1}, which builds upon the steady
state vertical structure calculations (Z2009a; an analytical analysis is
given in the Appendix).  The shaded regions in Figure \ref{fig:m1} are
the radii at which disk accretion is expected to be unsteady if the mass
infall rate is constant.

The solid curve farthest to the lower right in the Figure (labeled
$R_Q$) is the radius where the central temperature of a pure GI-driven
disk (Q=1 disk) would reach $T_M$ (e.g. 1400 K).  In other words, at a
given infall rate $\dot{M_{in}}$, pure GI disks can accrete steadily
beyond $R_{Q}$ by the GI, but within $R_{Q}$ the MRI will be activated.
Up and to the left of $R_Q$ in Figure \ref{fig:m1} another solid curve,
labeled $R_{M}$, denotes the radii at which a pure MRI disk of the given
$\alpha_M$ would have a central temperature of T$_{M}$. If the disk is
MRI active, it can accrete steadily purely by the MRI within $R_{M}$,
but beyond $R_{M}$ the MRI will be turned off.  When $R_M$ and $R_Q$
cross a smooth transition between the GI and MRI exists and steady
accretion is possible.  From the left panel of Figure \ref{fig:m1}, we see that
the disk can accrete steadily with $\alpha$=0.01 and
$\dot{M_{in}}=10^{-4}\msunyr$.  This is observed in the time
evolution discussed in \S 4.1
 and Figure \ref{fig:1em4} (see also \cite{armitage01}).

We predict that in the shaded regions matter will pile up through the
action of GI, trigger the MRI, and produce an outburst.  The dotted
curve shows $R_{Q}$ and $R_{M}$ if T$_{M}$=1800 K (at which temperature
all dust has sublimated). $R_{Q}$ and $R_{M}$ at 1800 K are smaller than
they are at 1400 K because of the temperature plateau around the dust
sublimation temperature (Z2009a; also can be seen at R$\sim$ 2-10 AU in
Figs. \ref{fig:14001800}). Thus if $T_M$ is higher, outbursts are
expected to be shorter because the outburst drains the small inner disk
($R < R_{Q}$) on a shorter viscous timescale (as demonstrated in Figures
\ref{fig:1em4}-\ref{fig:1em62}). The effect of $\alpha_{M}$ can also be
seen by comparing the left and right panels of Figure \ref{fig:m1}.

The classical thermal instability will also be triggered at the
infall phase as shown by the upper left shaded band in Figure
\ref{fig:m1}. Even the `steady'
  accretion case discussed above with
$\alpha$=0.01 and $\dot{M_{in}}=10^{-4}\msunyr$ is subject to
thermal instability at $R < 1 AU$.  The two lines shown in Figure
\ref{fig:m1} correspond to the two limiting values of the ``S
curve'' (e.g., Faulkner, Lin, \& Papaloizou 1983) at which
transitions
  to the high (rapid accretion) state and the low (slow
accretion) state occur.  The instability depends on the disk's
vertical structure (different ``S curves"
 ) which behaves
  differently
in 2-D than in 1-D, as discussed in \S 3.  We expect nature to
behave somewhat more like the 2-D than the
  1-D case, so for the steady
accretion model discussed above convection will add small variations
in $\dot{M}$ and
  we refer to these steady accretion cases as
quasi-steady. Generally, the thermally unstable region is
distinctive at $\mdot > 10^{-5} \msunyr$, which suggests the TI may
be common for protostellar disks.

The solid dots in Figure \ref{fig:m1} are the MRI trigger radii from
our 1-D simulations when $T_M = 1400$K.  Although the trigger radii
do not fall precisely on R$_{Q}$, most of them are in the shaded
region, indicating that this $\mdot - R$ plane has predicted
non-steady accretion, with potential outbursts, to occur for infall
rates $\lesssim 10^{-5} \msunyr$ for $\alpha_M = 0.01$ and $\lesssim
10^{-4} \msunyr$ for $\alpha_M = 0.1$. For $\mdot = 10^{-4} \msunyr$
and $\alpha_{M}=0.01$, the MRI-GI instability won't occur based on
this $\mdot - R$ plane, which agrees with the 1-D simulations.
Generally, our $\mdot - R$ plane results provide a good guide to the
parameters for which unsteady accretion occurs.

The outbursts and TI in
  quasi-steady accretion could provide
a much hotter thermal history for the protostellar disks, which may
have imprints on meteorites and the chemicals in the protostellar
disks.

\subsection{Steady vs Non-steady accretion}

The non-steady disk accretion in our model is the result of the
inability of the inner disk to transport mass inward at the same
rate as mass is fed in from the outer disk by infall,
$10^{-5}\msunyr$. This mass pileup eventually leads to MRI
activation and outbursts.  Even after infall ends, the dead zone
evolves with time if the mass addition from the outer disk due to GI
and viscous stresses exceeds what can be carried inward by the
active layer.

Terquem (2008) was able to construct steady-state disk accretion
solutions with active layers and "dead" zones.  The difference is
due to the assumptions of both finite, non-GI viscosity in the dead
zone and to assuming a very much lower mass accretion rate, $10^{-8}
\msunyr$.  At such low accretion rates, even very low non-GI dead
zone viscosities and comparable active layer properties suffice to
transport mass at these rates through the inner disk, which is not
possible in our disks driven on the outside at high mass infall
rates (except for small centrifugal radii, in which case the disk
becomes fully MRI-active).

\subsection{Constraints from observations of FU Orionis objects}

FU Orionis objects are outbursting systems with
  maximum disk accretion
rates $\dot{M}\sim$ 10$^{-4}\msunyr$ and a decay time of decades to
hundreds of years (Hartmann \& Kenyon 1996).  These properties constrain
our parameter space.

The outbursts produced by 1D2Z models are sensitive to $\alpha_{M}$
and T$_{M}$. Figures 2-4 show that $\alpha_{M}$=0.1, T$_{M}$=1800 K
leads to outbursts that are too strong ($\dot{M}$=10$^{-3}\msunyr$),
while $\alpha_{M}$=0.01, T$_{M}$=1400 K leads to outbursts that are
too long (3000 yr) \footnote{The observed decay time is
$\sim$$\tau_d = (d\ln F/dt)^{-1}$
($F$ is the flux; thus $\tau_d$ is an e-folding time
 ).  In principle the
outburst could have nonexponential time dependence and the duration of
the outburst could differ from $\tau_d$.  In our models, however, the
luminosity exceeds the preoutburst luminosity for a time only slightly
longer than $\tau_d$.}.  If T$_{M}$=1400 K, the outburst start at
$\sim$3 AU (Figure \ref{fig:14001800}), so $\alpha_M$ needs to be large
to produce the correct decay timescale.  If $T_M = 1800$K the outburst
is triggered at $\sim 1$AU, so $\alpha_M \sim 0.01$ is required 
 to maintain the
correct outburst timescale.

One uncertainty here is that $\alpha_{M}$ in a non-thermally ionized
active layer may not be the same as $\alpha_{M}$ in a thermally ionized,
dust free, fully conducting medium. Thus we mostly constrain
$\alpha_{M}$ in the latter (outburst stage) from the decay timescale;
$\alpha_{M}$ in the active layer is not well-constrained by outbursts.

Another uncertainty is related to our assumption of constant
$\Sigma_{A}$.  If $\Sigma_{A}$ is instead a function of radius the
disk's long term evolution will change.  At early stages, however,
if the GI in the outer disk dominates the disk's accretion,
variation of $\Sigma_{A}$ will have little effect on the outburst
mechanism discussed here.

The central star mass and infall rate do not significantly change the
outbursts. This agrees with the observation that FU Orionis outbursts
occur for protostars with different infall rates (Quanz et al. 2007 and
Zhu et al 2008 found that FU Orionis objects could be either Class 0 or
Class I objects.).

Zhu et al. (2007) argued that the decay timescale for FU Ori implies
$\alpha$ $\sim$ 0.02-0.2.  This conclusion can be tested with our 1-D
simulations. The last column in Table 1 shows the viscous timescale
calculated using $R_{M}$ and $T_{M}$:
\begin{equation}
t_{v}\sim \frac{R_{M}^{2}}{\nu}\,,
\end{equation}
where
\begin{equation}
\nu=\alpha \frac{c_{s}^{2}}{\Omega}\,,
\end{equation}
$c_{s}$ and $\Omega$ are calculated with $T_{M}$ and $R_{M}$.

Comparing with the outbursts' duration in column 6, we see the viscous
time is close to the outburst time for T$_{M}$=1400 K.  If T$_{M}$=1800
K, the viscous time is 2-3 times longer than the outburst time, which
may be due to the disk's temperature during the outburst being 
 higher than
T$_{M}$ if the MRI is triggered at smaller radius as T$_{M}$=1800 K
case. However, since constraining $\alpha$ by using the viscous
timescale is an
  order of magnitude estimate, our simulations are consistent
with the Zhu et al (2007) estimate.

\section{Conclusions}

In this paper, we have evolved a one-dimensional layered disk model
including both MRI and GI to study the unsteady disk accretion
of protostars. The 1-D models reproduce the general properties of
2-D (axisymmetric) outbursts reasonably well, such as the outbursting mass
accretion rate, duration
  and the accreted mass during one
outburst.  Because the 1-D model is faster, it
 enables us to study
outbursts in an extended parameter space.

Our results confirm that the disk can accrete steadily with high infall
rates ($\dot{M_{in}}$$\sim$10$^{-4}$$\msunyr$ if $\alpha_{M}$=0.01;
\cite{armitage01}).  This steady accretion may still have short
timescale variations, however, due to the thermal instability in the
inner disk, as suggested by our earlier, 2-D simulations.

We also confirm the prediction by Z2009a that protostars are likely to
accrete unsteadily/in outbursts for infall rates $\lesssim 10^{-5}
\msunyr$ with $\alpha_M = 0.01$ and $\lesssim 10^{-4} \msunyr$ for
$\alpha_M = 0.1$. Outbursts are triggered at $r \sim 1-10$~AU for
protostellar infall rates $\sim 10^{-5} - 10^{-6} \msunyr$. The
outbursts are stronger and shorter with larger $\alpha_{M}$ or T$_{M}$.
The total mass accreted during one outburst mainly depends on T$_{M}$.
While the outbursts are slightly shorter for more massive central stars,
the outburst $\dot{M}$ is nearly independent of central star mass.  The
active layer surface density only affects the mass accretion rate in
  the
low state; it has little effect on the outburst.

By comparing with 
 the mass accretion rate and duration of
  observed FU
Orionis events, we can constrain a combination of $\alpha_{M}$ and
T$_{M}$. If $\alpha_{M}$ is low (0.01), T$_{M}$ needs to be high
(1800 K, higher than the dust sublimation temperature); if
$\alpha_M$ is high (0.1) then $T_M$ needs to be low ($\lesssim$1400
K).

Our results show that 1-D, two zone models can capture the basic
features of disk evolution, given our assumptions about the action of
the MRI and GI.  In a later paper we will address disk evolution over a
much longer timescale, explicitly taking into account mass infall from a
rotating protostellar cloud.

\acknowledgments

This work was supported in part by NASA grant NNX08A139G, by the
University of Michigan, by a Sony Faculty Fellowship, a Richard and
Margaret Romano Professorial Scholarship, and a University Scholar
appointment to Charles Gammie.

\appendix

\begin{figure}
\includegraphics[width=0.42\textwidth]{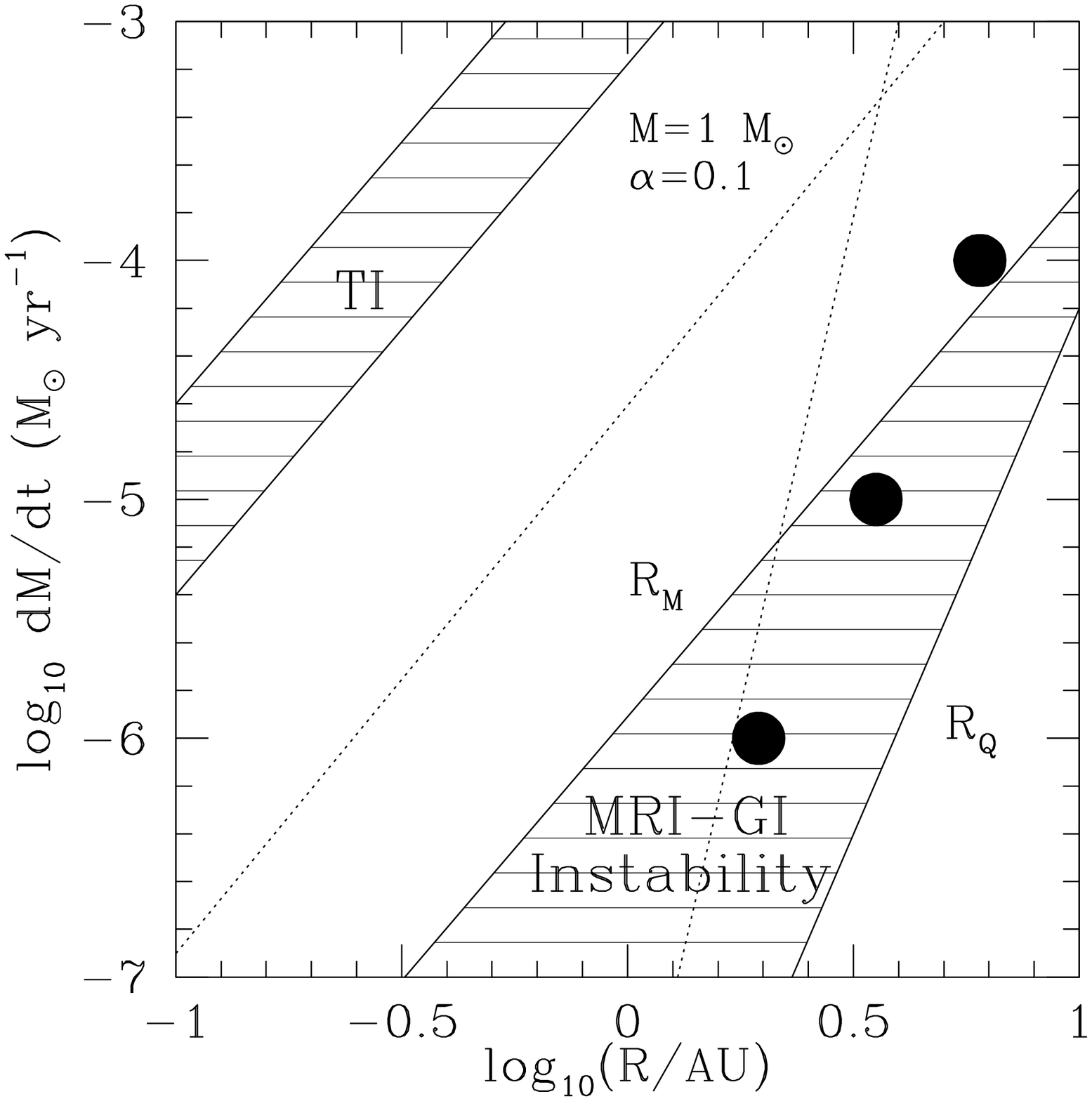} \hfil
\includegraphics[width=0.42\textwidth]{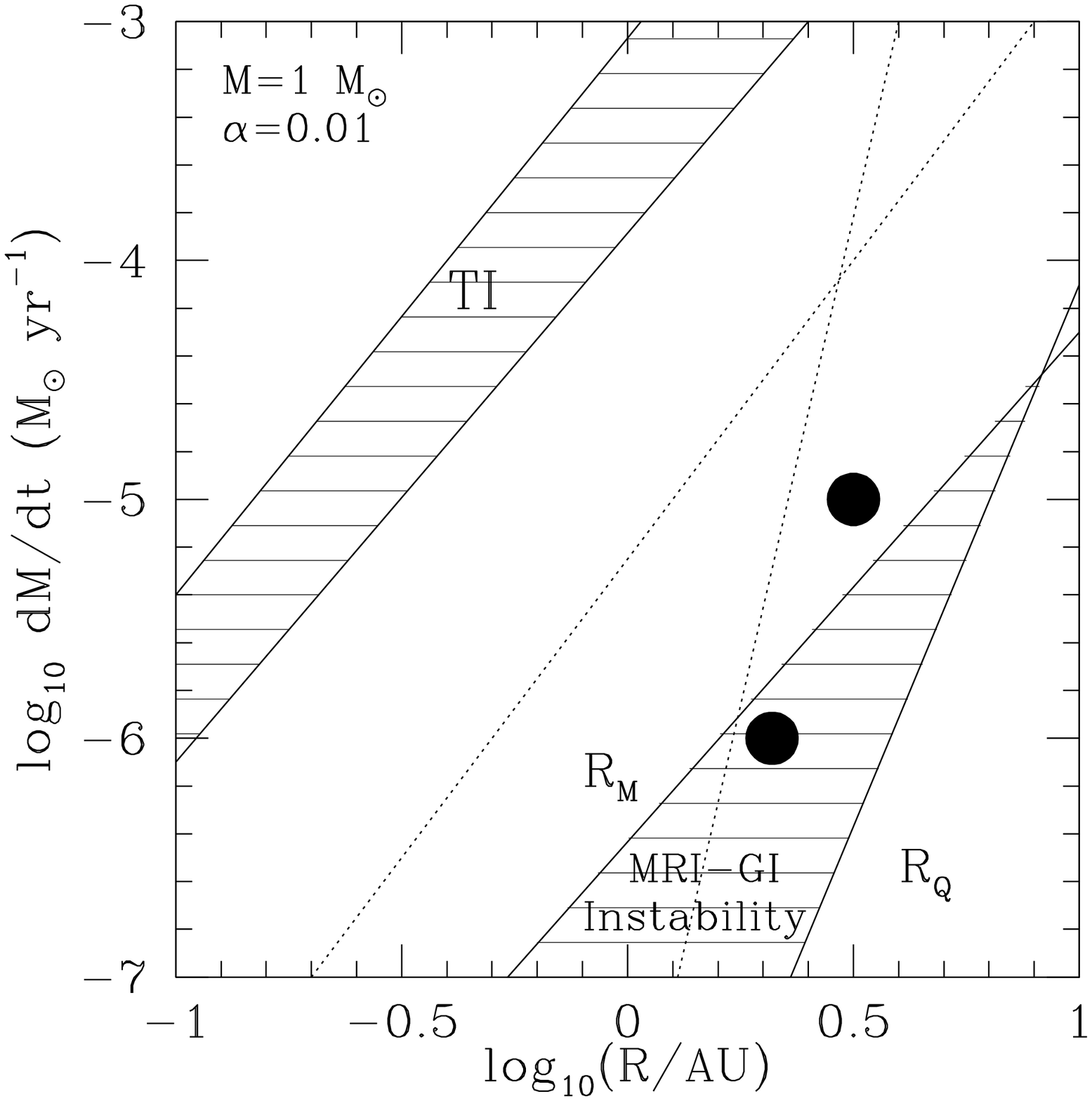} \\
\caption{ Unstable regions in the $R-\dot{M}$ plane for a $1 \msun$
central star. The shaded region in the lower right shows the MRI-GI
instability with the MRI trigger temperature of $1400$ K.  The
dotted curves show $R_M$ and $R_Q$ (the boundaries of the MRI-GI
instability shaded region; see text for definition) for an MRI
trigger temperature of $1800$ K. The shaded region in the upper left
shows the region subject to classical thermal instability. The solid
dots represent the radii where the MRI is triggered in 1-D
simulations for $T_{M}$=1400 K.} \label{fig:m1}
\end{figure}

By assuming a marginally gravitationally stable (Q=1.5) disk, the
disk's structure is determined with a given $\dot{M}$, and thus the
radius where the outburst is triggered (R$_{Q}$ in Fig. \ref{fig:m1}
and Z2009a) can be derived. Unlike Z2009a where the detailed
vertical structure is calculated numerically to give R$_{Q}$, here
we give simple analytical results for R$_{Q}$ by assuming the disk
is vertically isothermal with constant opacity at a given radius.

First, if $\kappa$=CT$^{\alpha}$P$^{\beta}$, the relationship
between $\Sigma$ and the central (midplane) temperature T$_{c}$ is
given by
\begin{equation}
T_{c}^{4}=\frac{3}{8}T_{eff}^{4}\tau=\frac{3}{16}T_{eff}^{4}\Sigma\kappa\,.
\end{equation}
Using the form for $\kappa$ and using
   $\rho_{c}$=$\Sigma/2H$, where H
is the disk scale height,
\begin{eqnarray}
T_{c}&=&3^{1/(4-\alpha-\beta/2)}2^{(-4-\beta)/(4-\alpha-\beta/2)}T_{eff}^{4/(4-\alpha-\beta/2)}\Sigma^{(1+\beta)/(4-\alpha-\beta/2)}C^{1/(4-\alpha-\beta/2)}\nonumber\\
&&\times\Omega^{\beta/(4-\alpha-\beta/2)}\left(\frac{k}{ \mu
m_{H}}\right)^{\beta/(8-2\alpha-\beta)}\,,
\end{eqnarray}
or equivalently
\begin{equation}
\Sigma=3^{-1/(1+\beta)}2^{(4+\beta)/(1+\beta)}T_{eff}^{-4/(1+\beta)}T_{c}^{(4-\alpha-\beta/2)/(1+\beta)}C^{-1/(1+\beta)}\Omega^{-\beta/(1+\beta)}\left(\frac{k}{
\mu m_{H}}\right)^{-\beta/(2+2\beta)}\,,\label{eq:sigma}
\end{equation}
 where $\Omega$ is the angular velocity at R,
$k$ is the Boltzmann constant and m$_{H}$ is the unit molecular
mass.

Then with Q=c$_{s}$$\Omega$/$\pi$G$\Sigma$, and inserting equation
(\ref{eq:sigma}) into Q 
   to derive the relationship between R and
$T_{c}$ at a given Q and $\dot{M}$
\begin{eqnarray}
  R&=&3^{4/(9+6\beta)}2^{(-14-2\beta)/(9+6\beta)}\pi^{(-4-2\beta)/(9+6\beta)}\left(\frac{\dot{M_{in}}}{
\sigma}\right)^{2/(9+6\beta)}T_{c}^{(2\alpha+2\beta-7)/(9+6\beta)}C^{2/(9+6\beta)} \nonumber\\
&&\times G^{1/(9+6\beta)}M^{(3+2\beta)/(9+6\beta)}\left(\frac{k}{
\mu
m_{H}}\right)^{(1+2\beta)/(9+6\beta)}Q^{(-2-2\beta)/(9+6\beta)}\,.\label{eq:R}
\end{eqnarray}

The dust opacity fitting from Z2009a suggests, at T$\lesssim 1400$
K, $\alpha$=0.738, $\beta$=0 and C=0.053. If we plot the
relationship between R and $\dot{M}$ by given T$_{c}$=T$_{M}$=1400 K
and Q=1.5, we find
\begin{equation}
R_{Q}=11
\left(\frac{\dot{M_{in}}}{10^{-4}\msunyr}\right)^{2/9}\left(\frac{M}{M_{\odot}}\right)^{1/3}AU\,,
\end{equation}
which corresponds well with the R$_{Q}$ calculated in Figure \ref{fig:m1}.

Since R$_{Q}$$\propto$$\dot{M}^{2/9}$ for $T_{M}$=1400 K case, the
outburst timescale is roughly
\begin{equation}
t\sim\frac{R_{Q}^{2}}{\nu}\sim 960\frac{0.1}{\alpha_{M}}
\left(\frac{\dot{M_{in}}}{10^{-4}\msunyr}\right)^{1/9}\left(\frac{M}{M_{\odot}}\right)^{2/3}
\rm{yr}\,,
\end{equation}
where $\nu$ is calculated for T$_{c}$=T$_{M}$=1400 K. During
outburst, however, T$_{c}$ could be higher than the MRI trigger
temperature T$_{M}$, especially in the inner part of the disk.

If we further assume the outburst $\dot{M}\sim\nu\Sigma(R_{Q})$
(this is the steady accretion disk solution which may not be true in
the time-dependent case), we find
\begin{equation}
\dot{M}\sim3\pi\nu\Sigma\sim\frac{3\alpha_{M} c_{so}^{2}c_{st}}{ G
Q}\,.
\end{equation}
Here we explicitly distinguish between the sound speed c$_{so}$
during the outburst and the sound speed c$_{st}$ before MRI
activation.  For an order of magnitude estimate we assume $c_{so}
\sim c_{st} \sim c_s(1400 {\rm K})$ and so
\begin{equation}
\dot{M}=5\times10^{-4}\; \frac{\alpha_{M}}{0.1}\; \msunyr
\end{equation}
which agrees with the numerical simulations for T$_{M}$=1400 K
reasonably well considering we are using a steady state assumption.

We have only applied the above equations for the
T$_{M}$=1400 K case, since the detailed vertical structure is
important if T$_{M}$=1800 K where the midplane is dust-free while
the surface has dust. Also the
outburst is triggered at a 
 smaller radius for T$_{M}$=1800 K, and thus
the disk temperature during the outburst can be much higher than the
MRI trigger temperature.

\clearpage

\begin{table}
\begin{center}
\caption{1D2Z models \label{tab1}}
\begin{tabular}{ccccccccc}

\tableline\tableline
M$_{*} $  & infall rate &$\alpha$ & T$_{M} $ & outburst $\dot{M}$\tablenotemark{a} & duration & accreted mass\tablenotemark{b} & R$_{M}$\tablenotemark{c} & viscous time\tablenotemark{d}\\
M$_{\odot}$& $\msunyr$ & & K & $\msunyr$ & yr & M$_{\odot}$ &AU & yr\\
\tableline
1 & 10$^{-4}$ & 0.1 & 1400 & 2$\times$10$^{-4}$ & 700 & 0.7 & 6 & 428 \\
1 & 10$^{-4}$ & 0.1 & 1800 & 10$^{-3}$ & 50 & 0.02 & 1.8 & 182\\
1 & 10$^{-5}$ & 0.1 & 1400 & 2$\times$10$^{-4}$ & 400 & 0.057 & 3.5 & 327\\
1 & 10$^{-5}$ & 0.1 & 1800 & 10$^{-3}$ & 40 & 0.024 & 1.2 & 150 \\
1 & 10$^{-5}$ & 0.01 & 1400 & 5$\times$10$^{-5}$ & 4000 & 0.045 & 3.2 & 3127\\
1 & 10$^{-5}$ & 0.01 & 1800 & 8$\times$10$^{-5}$ & 400 &0.01 & 1.2 & 1490 \\
1 & 10$^{-6}$ & 0.1 & 1400 & 2$\times$10$^{-4}$ & 350 & 0.08 &1.9 & 240 \\
1 & 10$^{-6}$ & 0.1 & 1800 & 10$^{-3}$ & 50 & 0.02 & 0.85 & 125\\
1 & 10$^{-6}$ & 0.01 & 1400 & 5$\times$10$^{-5}$ & 3000 & 0.04 & 2.1 & 2533\\
1 & 10$^{-6}$ & 0.01 & 1800 & 8$\times$10$^{-5}$ & 800 & 0.015 & 0.87 & 1268\\
0.3 & 10$^{-5}$ & 0.1 & 1400 & 2$\times$10$^{-4}$ & 300 & 0.04 & 2.3 & 145 \\
0.1 & 10$^{-5}$ & 0.1 & 1400 & 2$\times$10$^{-4}$ & 250 & 0.025 & 1.43 & 66\\
 \tableline
\end{tabular}
\tablenotetext{a}{the mass accretion rate at the half-time of the
outburst}
\tablenotetext{b}{the mass accreted during one outburst}
\tablenotetext{c}{the MRI trigger radius}
 \tablenotetext{d}{The viscous timescale is calculated by using R$_{M}$ and T$_{M}$.}
\end{center}
\end{table}
\clearpage

\FloatBarrier

\end{document}